\documentclass[aps,pra,preprint,superscriptaddress]{revtex4-1}
\usepackage{graphicx}
\usepackage{amssymb}
\usepackage{amsmath}
\usepackage{amsfonts}
\usepackage{graphicx}
\usepackage{soul}
\usepackage{float}

\usepackage{color}

\makeatletter
\newcommand*{\rom}[1]{\expandafter\@slowromancap\romannumeral #1@}
\makeatother

\begin{document}


\title{Localized patterns in star networks of Class B lasers with optoelectronic feedback}


\author{J. Shena}
\author{J. Hizanidis}
\email[]{hizanidis@physics.uoc.gr}
\affiliation{Department of Physics, University of Crete, 71003 Heraklion, Greece}
\affiliation{National University of Science and Technology MISiS, Leninsky prosp. 4, Moscow, 119049, Russia}
\author{N. E. Kouvaris}
\affiliation{Department of Mathematics, Namur Institute for Complex Systems (naXys), University of Namur, Rempart de la Vierge 8, B 5000 Namur, Belgium}
\author{G.~P. Tsironis}
\affiliation{Department of Physics, University of Crete, 71003 Heraklion, Greece}
\affiliation{National University of Science and Technology MISiS, Leninsky prosp. 4, Moscow, 119049, Russia}


\date{\today}

\begin{abstract}
We analyze how a star network topology shapes the dynamics of coupled CO\textsubscript{2} lasers with an intracavity electro-optic modulator that exhibit bistability.
Such a network supports spreading and stationary activation patterns.
In particular,
we observe an activation spreading where the activated periphery turns on the center element, an activated center which drifts the periphery into the active region and an activation of the whole system from the passive into the active region. 
Pinned activation, namely activation localized only in the center or the peripheral elements is also found.
Similar dynamical behavior has been observed recently in complex networks of coupled bistable chemical reactions.
The current work aims at revealing those phenomena in laser arrays, giving emphasis on the essential role of the coupling structure in fashioning the overall dynamics.
\end{abstract}


\maketitle

\section{Introduction}
\label{sec:introduction}

Solid-state and semiconductor laser arrays constitute a wide family of nonlinear coupled systems with complex dynamic behavior. 
Although the emission from the individual units is often unstable with large amplitude chaotic pulsations \cite{Wang1988, Fabiny1993, Thornburg1997}, the coupled system can show synchronization and other spatiotemporal phenomena \cite{Rogister2007}. 
The main difference between semiconductor and solid-state lasing media lies in the value of the linewidth enhancement factor $a$, which is $3 \leq a \leq 5$ for semiconductor and $a=0$ for solid-state systems.
This difference makes solid-state lasers more suitable in applications where phase locking is required.

In recent years, there have been many studies concerning semiconductor lasers and the analysis of synchronization and chimera states \cite{Winful1992,Bohm2015,Shena2017,Shena2017a}.
Here though, we focus on solid-state laser arrays and the formation of localized stationary patterns of activity.
The dynamic behavior of each laser element is bistable and the coupling between the elements is local and arises due to the overlap of the electric fields of each separated beam \cite{Zehnle2000, Fabiny1993}. 
The theoretical model we use is originated from numerical and experimental studies of a CO\textsubscript{2} laser with an intracavity electro-optic modulator that exhibits bistability \cite{Wang1990}.
This model has many similarities to that obtained by semiconductor lasers with a saturable absorber inside the cavity \cite{Yamada1993}.
A similar problem was revisited for a Nd:YAG laser with an acousto-optic modulator \cite{Meucci2002}. 
Bistability were found in semiconductor lasers with strong optical injection \cite{Wieczorek2002} and in semiconductor laser diodes with a saturable absorber \cite{Dubbeldam1999}.

Other classical examples where bistable behavior is encountered are dynamical processes in chemical systems \cite{Mikhailov1990, Epstein1998}.
Recently, studies on complex networks of coupled bistable chemical reactions revealed rich collective dynamics, such as spreading or retreating of an initial activation, but more interestingly, the formation of localized stationary patterns dependent on the coupling strength and the degree distribution of the nodes \cite{Kouvaris2012, Kouvaris2013, Kouvaris2016, Kouvaris2017}.
Beyond the simplified theoretical approach, electrochemical experiments \cite{Kouvaris2016, Kouvaris2017} have stressed that the coupling topology plays a significant role in the observed dynamics resulting in a robust pattern formation mechanism.
Therefore, similar findings are expected to be seen in laser arrays coupled in such a way thus forming complex networks.

Here we focus on the simple case of star networks where each bistable element is connected to a central one, the hub.
This connectivity structure is often found in many natural or engineered systems that consist of dynamical elements interacting with each other through a common medium. 
It has also been used in optically coupled semiconductor lasers \cite{Zamora-munt2010, Bourmpos2012} where synchronization phenomena were investigated. 
We present an extended numerical analysis that takes advantage of the simplicity of the star network topology to determine the conditions required for the formation of localized stationary patterns. 
We start our investigation by analyzing the dynamics and determining the bistable regime for a single laser. 
In the bistable regime the active and the passive states of the laser coexist.
Knowing this we explore the dynamics of two coupled bistable lasers, before we proceed to our main study for a star network of such elements.
This work comments on the formation mechanism of stationary patterns which --like in the electrochemical networks-- is strongly dependent on the role of the coupling topology.

\section{The Model}
\label{model}

The dynamical behavior of the CO\textsubscript{2} laser with feedback can be described by three coupled first-order differential equations, one for the laser field ($E$), the second for the population inversion ($G$) and the last for the feedback voltage of the electro-optic modulator ($V$). 
In dimensionless form, the evolution equations have the form \cite{Wang1990}

\begin{subequations}
\begin{eqnarray}
\frac{dE}{dt} &=& \frac{1}{2} \left (G-1-a\sin^{2}(V) \right ) E  \\ 
\frac{dG}{dt} &=&\gamma(P-G-G \vert E \vert^{2} ) \\
\frac{dV}{dt} &=&\beta(B+f\vert E\vert ^{2}-V)\,,
\end{eqnarray}\label{eq1}
\end{subequations}

\noindent where $\vert E \vert$ is the amplitude of electric field,
$\gamma$ denotes the population decay time, $P$ denotes the pumping and $a$ scales the maximum loss introduced by the modulator. 
The damping rate $\beta$ of the feedback loop is normalized to the cavity decay rate, $B$ is the bias voltage applied to the modulator amplifier, and $f$ is the scaling of the feedback gain, i.~e. it measures the relation between the intensity incident on the photodiode and the voltage delivered by the differential amplifier. 
In general, $B$ is used as a control parameter.

In the case of a single laser, the phase of the electric field is a constant variable in time and has no role in the system dynamics \cite{ThomasErneux}. 
Thus, we prefer to work with the amplitude of the electric field without loss of generality. 
In this framework, the system of Eqs.~\eqref{eq1} admits the zero-intensity solution $(\vert E \vert=0,G=P,V=B)$ and the non-zero intensity solution(s) which are given in the parametric form

\begin{subequations}
\begin{eqnarray} 
\frac{P}{1+ \vert E \vert ^{2}} &=& 1 +a\sin^{2}\left( B+f \vert E \vert ^{2}\right) \\ 
G &=& \frac{P}{1+\vert E \vert^{2}} \nonumber \\ 
V &=& B+f\vert E \vert^{2}\,.
\end{eqnarray}\label{eq2}
\end{subequations}

\begin{figure}[t]
\includegraphics[scale=0.5]{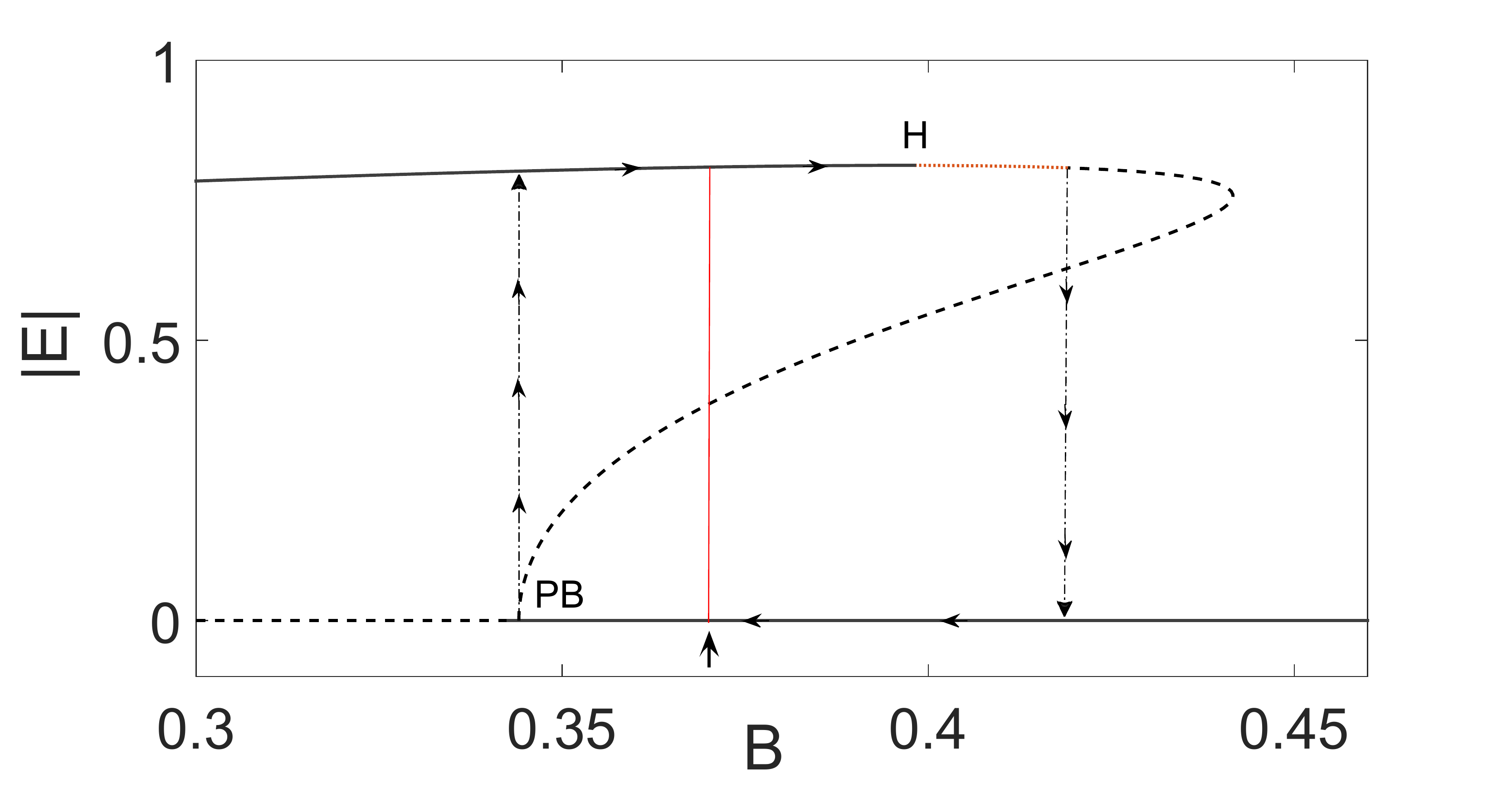}
\caption{High-gain bifurcation diagram. The stationary amplitude of the laser field $\vert E \vert$ is shown as a funtion of the bias voltage $B$. 
The solid and dashed lines mark the stable and unstable branches, respectively, while the arrows indicate the hysteresis loop. 
H denotes the Hopf bifurcation point and PB the subcritical pitchfork bifurcation. 
The constant value of $B=0.37$ that will be used in the following sections has been indicated by the arrow and the red line. 
Other parameters are $\gamma = 0.003125$, $P = 1.66$, $\beta = 0.0521$, $ a = 5.8$, and $f = -0.6$.
Dotted line denotes a very small regime of low-amplitude oscillations.}
\label{fig_1}
\end{figure}

Figure ~\ref{fig_1} illustrates the stability of these fixed points by studying the bifurcation diagram in the case of high gain $f=-0.6$ and using $B$ as the control parameter \cite{Wang1990}. 

For $B<0.3441$, the zero-intensity fixed point is unstable (marked by the dashed line), and the nonzero-intensity solution is the only attractor in the system (active state).
At $B=0.3441$, a subcritical pitchfork bifurcation (PB) takes place and the zero-intensity fixed point becomes stable. 
At the same time, a new unstable fixed point is born which vanishes for $B=0.3982$.
In the interval $0.3441 < B < 0.4191$ the system exhibits bistability and a hysteresis loop is observed. 
As $B$ increases beyond the value $0.4191$, the zero intensity solution is the only allowed state in the system. 
In the rest of our analysis we hold the bias voltage constant and equal to $B=0.37$
in order to achieve a controllable bistable system that can be prepared either in the passive state $0<\vert E \vert<0.2$ or the active one $0.7< \vert E \vert<0.9$. 
Moreover, the chosen value $B=0.37$ allows us to avoid transitions from the Hopf point at $B=0.3982$, above which a very small regime of low-amplitude 
oscillations exists (marked by the dotted line). 
The bifurcation diagrams throughout the manuscript have been generated using the MatCont software, a numerical continuation package for the interactive bifurcation analysis of dynamical systems \cite{Govaerts2011}.

\section{Two coupled bistable lasers}

\begin{figure}[t]
\includegraphics[scale=1.5]{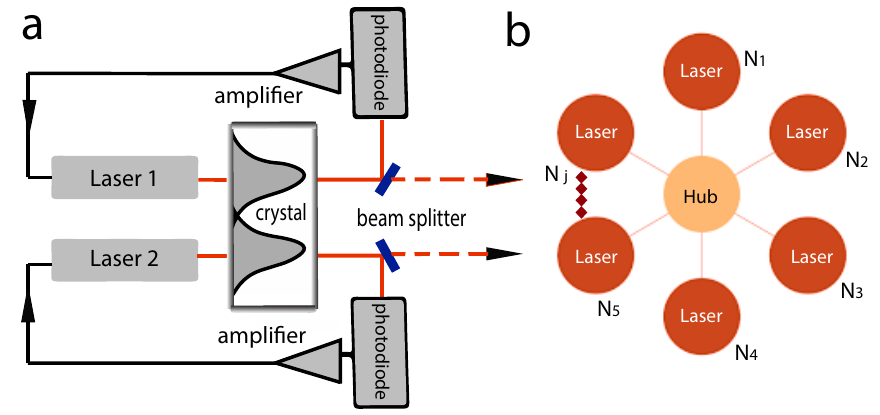}
\caption{a) Schematic diagram of the optoelectronic feedback of two coupled lasers. 
The optical power emitted by the two lasers is coupled through the overlap
of the electrical fields in a nonlinear crystal. After a beam splitter, it is detected by a photodiode with a fixed bandwidth. 
The electrical output is fed back to each laser through an amplifier. 
b) Topology of a star network where each laser of the periphery interacts with the rest
through a central laser, the hub, with tha same coupling strength. \label{fig_2}}
\end{figure}

Having defined the single bistable laser system, we proceed by considering two parallel waveguides of CO\textsubscript{2} lasers each one with a proper optoelectronic feedback (see Fig. \ref{fig_2}).
The mutual interaction lies on the overlap integrals of both lasers fields inside the crystal with a proper refractive index profile \cite{Zehnle2000}. 
The evolution equations for this coupled system have the form

\begin{subequations}
\begin{eqnarray}
\frac{dE}{dt} &=& \frac{E}{2}(G-1-a\sin^{2}(V)) + \eta E_{H}  \\ 
\frac{dE_{H}}{dt} &=& \frac{E_{H}}{2}(G_{H}-1-a\sin^{2}(V_{H}))+\eta E\,,
\end{eqnarray}
\label{eq3}
\end{subequations}

\noindent where the subscript H denotes the second laser. 
The equations for the population inversion ($G$ and $G_{H}$) and the feedback voltage of the modulator ($V$ and $V_{H}$) have the same form as in Eqs.~\ref{eq1}, therefore we omit them. 
The parameter $\eta$ is the coupling strength between the two lasers and in general is a complex parameter ($\eta=\eta_{\Re}+i\eta_{\Im}$). 
The real part $\eta_{\Re}$ takes usually negative values and vanishes only when $D \simeq 2w$, where $D$ is the distance between the two beams and $w$ is the waist of the beam Gaussian portrait.
However, it is possible to have positive coupling values, which we consider here, by
pumping in the middle between the two beams~\cite{Laabs1997}. 
The imaginary part $\eta_{\Im}$, is related to the refractive index and can be zero for a laser beam of weak intensity, which is the case here. 
If we use polar coordinates $E=\vert E \vert e^{i\phi}$, a third equation for the phase difference of the two lasers is added to Eqs.~\ref{eq3}. 
However, we can neglect the dynamics of the third variable since we are working in the phase locking regime, i.~e. the phase difference is constant and equal to zero (See Fig.~S1 in Supplemental Material (SM) \cite{SM}). The dynamics of the system can, therefore, be described solely by the amplitude of the electric field.

The zero intensity steady state of the coupled system is equal to $|E|=|E_{H}|=0,\ G=G_{H}=P,\ V=V_{H}=B$, while the non-zero intensity steady states of Eqs.~\ref{eq3}, are given in the parametric form

\begin{subequations}
\begin{eqnarray} 
\frac{P}{1+\vert E \vert^{2}} &=& 1 +a\sin^{2}\left( B+f \vert E \vert^{2}\right) + 2 \eta \frac{ \vert E_{H} \vert}{\vert E \vert} \\ 
\frac{P}{1+\vert E_{H} \vert^{2}} &=& 1 +a\sin^{2}\left( B+f \vert E_{H} \vert^{2}\right) + 2 \eta \frac{\vert E \vert}{\vert E_{H} \vert}\,.
\end{eqnarray}\label{eq4}
\end{subequations}

\begin{figure}[t]
\includegraphics[scale=0.5]{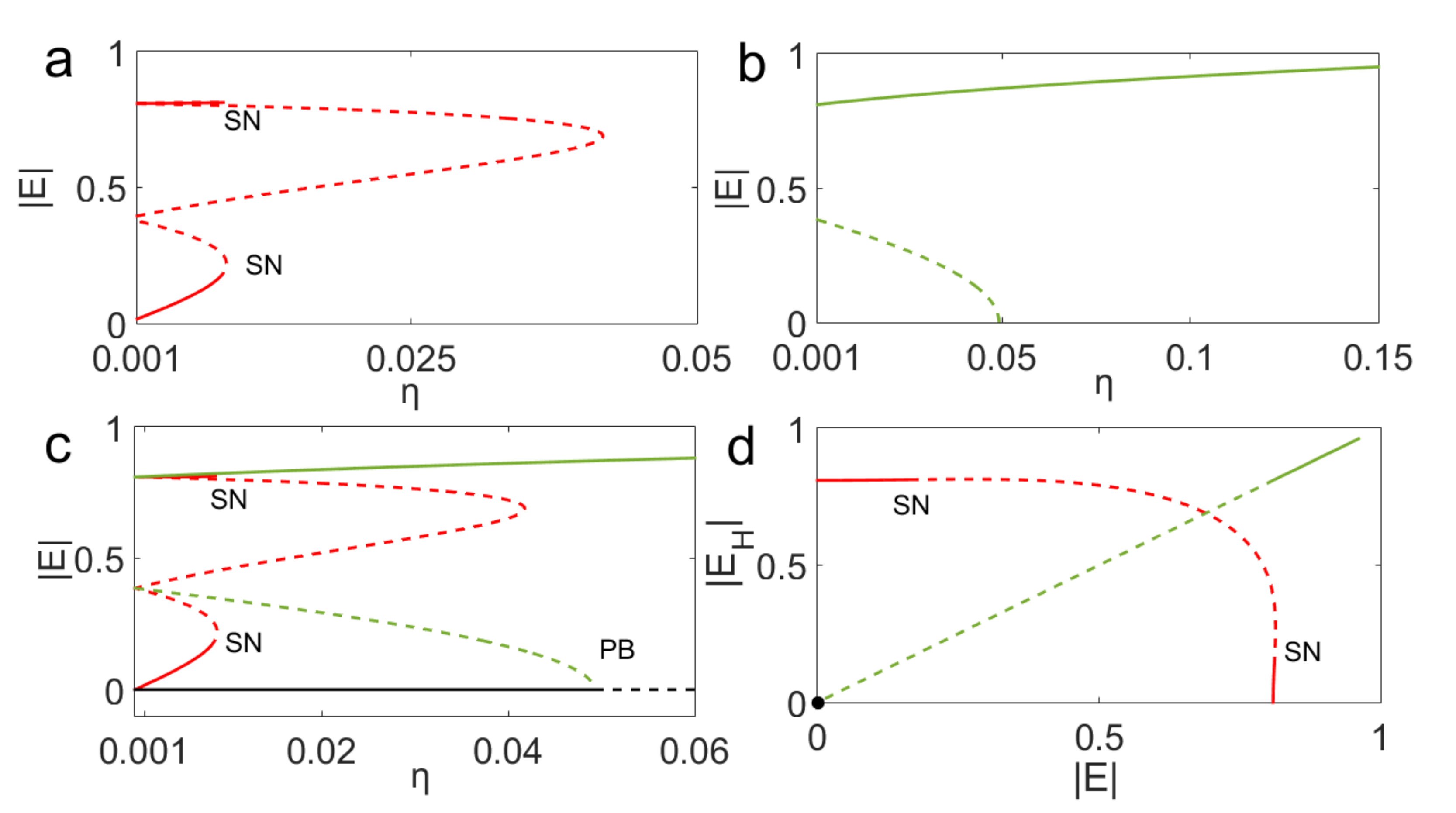}
\caption{The stationary amplitude of the laser field versus the coupling strength. a) The amplitude of the first laser in the case where the system is prepared with the first laser located in the passive state and the second in the active state (passive-active). 
b) The stationary amplitude of the laser field versus the coupling strength where both the two lasers are located in the active state (active-active).
c) The stability for all three preparation states of the system (passive-active, active-active and passive-passive). 
d) The stability region for the amplitude of both lasers in all three cases is shown in the $(\vert E \vert, \vert E_H \vert)$ plane.
Solid and dashed lines correspond to stable and unstable steady states, respectively. 
SN stands for the saddle-node bifurcation, while PB for the subcritical pitchfork. 
$B=0.37$ and all other parameters as in Fig. \ref{fig_1}.
\label{fig_3}}
\end{figure} 

Figure \ref{fig_3} shows the stability of the system steady states as a function of the coupling strength.
Figure~\ref{fig_3} (a) shows the stationary amplitude of the electric fields in the case where the system is prepared with the first laser in the passive and the second one in the active state. 
Similarly, Fig.~\ref{fig_3} (b) shows the stability of the system when both lasers are prepared in the active state.
In Fig.~\ref{fig_3} (c) the two previous cases (passive-active, active-active) are plotted together with the passive-passive state which corresponds to the black line.
Figure~\ref{fig_3} (d) shows the stability region for the amplitude of both lasers in all three cases is shown in the $(\vert E \vert, \vert E_H \vert)$ plane, with the passive-passive state represented by a full black circle.
The thick and dashed lines correspond to the stable and unstable solution branches, respectively.
From Fig.~\ref{fig_3} (c) we can see that the passive-passive state (black line) undergoes a subcritical pitchfork bifurcation (PB) at $\eta=0.04922$. 
The passive-active state (red line) is stable up to a critical coupling strength $\eta=0.008819$, where a saddle-node bifurcation (SN) occurs, and the active-active state (green line) is stable for the whole $\eta$ range and coexists with an unstable branch that emerges at the PB point and runs through the negative axis (not shown here because $\eta$ has physical meaning only for positive values).

As a result of the above stability analysis, when the system starts at the passive-active state, both lasers jump to the active state (green branch) through a SN bifurcation at rather low coupling strengths $\eta>0.008819$. 
This resembles the chemical bistable media where an activation front can propagate thus activating the passive nodes \cite{Kouvaris2016, Kouvaris2017}.
On the other hand, when the system is prepared in the passive-passive state, higher values $\eta>0.04922$ are required for the lasers to jump to the active-active state, and this transition takes place through a PB bifurcation.
This spontaneous activation arises for the actual nature of the coupled laser system and is not observed in chemical bistable media \cite{Kouvaris2016, Kouvaris2017}.
Finally, if the system is prepared in the active-active state (green line), both lasers remain there for all values of the coupling strength.

\section{Star network of coupled bistable lasers}
Having analyzed the dynamics of two coupled bistable lasers, we now focus on a star network configuration and how it contributes to the formation mechanism of stationary active patterns.
In such a system each element of the periphery interacts with the rest through a central element, the hub (see Fig,~\ref{fig_2} (b)), thus Eqs.~\eqref{eq3} can be reformulated as follows: 

\begin{subequations}
\begin{eqnarray}  
\frac{dE_{j}}{dt} &=& \frac{E_{j}}{2}(G_{j}-1-a\sin^{2}(V_{j})) + \eta E_{H}  \\ 
\frac{dE_{H}}{dt} &=& \frac{E_{H}}{2}(G_{H}-1-a\sin^{2}(V_{H}))+\eta \sum_{j=1}^{N} E_{j}\,,
\end{eqnarray}\label{eq5}
\end{subequations}

\noindent where $j=1,2 \dots N$ counts for the number $N$ of the peripheral elements and the subscript $H$ denotes the hub. 
In polar coordinates Eqs.~\ref{eq5} become:

\begin{subequations}
\begin{eqnarray} 
\frac{d \vert E_{j} \vert}{dt} &=& \frac{1}{2} \vert E_{j} \vert \left[G_{j}-1-a \sin^{2} \right(V_{j} \left) \right] +\eta \vert E_{H} \vert \cos(\theta_{j})\label{eq6a} \\ 
\frac{d \vert E_{H} \vert}{dt} &=& \frac{1}{2} \vert E_{H} \vert \left[G_{H}-1-a \sin^{2} \right(V_{H} \left) \right] +\eta \sum_{j=1}^{N} \vert E_{j} \vert \cos(\theta_{j})\label{eq6b} \\ 
\frac{d \theta_{j}}{dt} &=& -\eta \left[ \frac{\vert E_{H} \vert}{\vert E_{j} \vert} \sin(\theta_{j}) +  \sum_{k=1}^{N}  \frac{\vert E_{k} \vert}{\vert E_{H} \vert}  \sin(\theta_{k}) \right] \label{eq6c}\,,
\end{eqnarray}\label{eq6}
\end{subequations}

\noindent where $\theta_{j}=\phi_{H}-\phi_{j}$ are the phase differences between the electric fields of each node of the periphery and that of the hub. 
The equations for the variables $ G_{j},\ V_{j},\ G_{H},\ V_{H}$ have the same form with Eqs.~\eqref{eq1} and, again, we omit them.
Numerical integration of Eqs.~\eqref{eq6} shows that in the $N-\eta$ parameter space the phase differences $\theta_{j}$ remain constant and equal to zero for $\eta>0.002$ (see Fig.~S2 in the Supplemental Material (SM) \cite{SM}).
Therefore, Eq.~\eqref{eq6c} can be neglected, the cosine terms are equal to 1, and the index $j$ can be dropped, reducing the star network to a system of two coupled lasers with asymmetric coupling.

\begin{subequations}
\begin{eqnarray}
\frac{d\vert E \vert }{dt} &=& \frac{1}{2} \vert E \vert \left[D-1-a \sin^{2} \right(V \left) \right] +\eta \vert E_{H} \vert \\ 
\frac{d \vert E_{H} \vert}{dt} &=& \frac{1}{2} \vert E_{H} \vert \left[D_{H}-1-a \sin^{2} \right(V_{H} \left) \right] +\eta N \vert E \vert\,.
\end{eqnarray} \label{eq7}
\end{subequations}

\noindent Previous studies with electrochemical systems \cite{Kouvaris2016, Kouvaris2017, Wickramasinghe2011} have implemented similar methods for reducing star and tree networks to chains of asymmetrically coupled nodes. In those theoretical and experimental studies, it was demonstrated that such a reduced system could produce all the rich dynamics of the original network despite its simpler form.

\begin{figure}[t]
\includegraphics[scale=0.5]{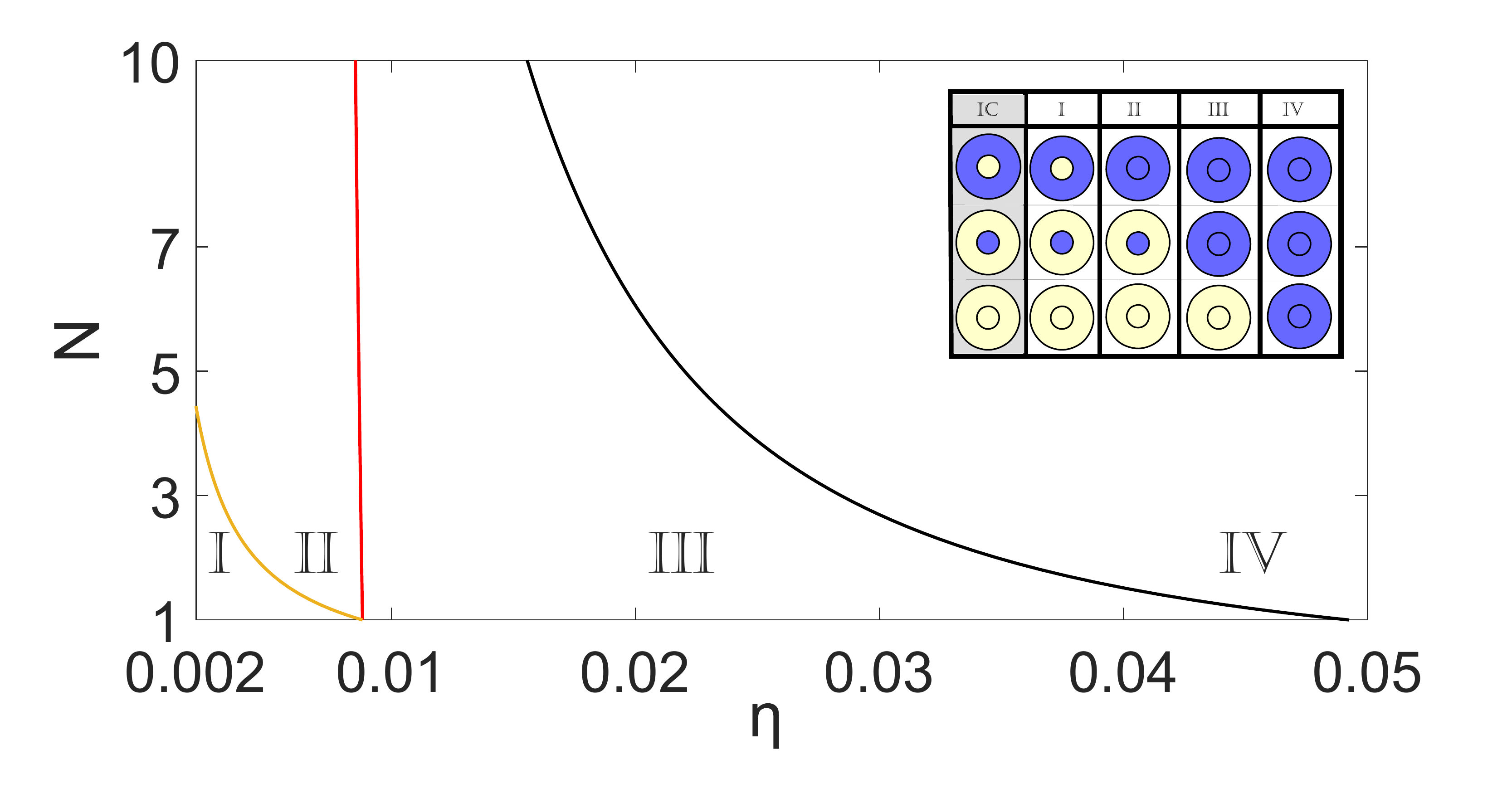}
\caption{Phase diagram in the $(\eta,N)$ parametric space.
Four dynamical regions are separated by curves that correspond to the continuation of the bifurcation points shown in Fig. \ref{fig_3} (orange and red curves correspond to saddle-node bifurcation lines, while the black curve corresponds to a pitchfork bifurcation line). 
In region \rom{1} the coupling is weak enough and all three initial conditions (IC) shown in the inset are stable and consist steady states of the system.
In region \rom{2} the active periphery drifts the hub to the active state. 
In region \rom{3} the active periphery drifts the hub to the active state but also the active hub drifts the periphery in the active state. 
In region \rom{4} the whole network goes to the active state.
In the inset the inner circle represents the hub and the outer circle represents the periphery, while the active state is denoted with blue color and the passive state with yellow.
Other parameters as in Fig. \ref{fig_3}. \label{fig_4}}
\end{figure}

Again, the zero intensity solution corresponds to $\vert E \vert=\vert E_H \vert=0,\ G=G_H=P,\ V=V_H=B$ and the non-zero intensity solutions are given in the parametric form

\begin{subequations}
\begin{eqnarray}
\frac{P}{1+\vert E \vert ^{2}} -1 - a \sin^{2} \left(  B+f \vert E \vert ^{2} \right)  &=&  -2 \eta \frac{\vert E_{H} \vert}{\vert E \vert} \\ 
\frac{P}{1+\vert E_{H} \vert ^{2}} -1 - a \sin^{2} \left(  B+f \varepsilon_{H}^{2} \right) &=& -2 \eta N \frac{\vert E \vert}{\vert E_{H} \vert}\,.
\end{eqnarray} \label{eq7}
\end{subequations}

In the previous section, the system of two symmetrically coupled lasers has shown that, depending on the preparation of the coupled system, the lasers transition to the active state occurs either through a saddle-node or a subcritical pitchfork bifurcation.
In order to locate these transitions in the star-network, we perform a continuation of the bifurcations in the $(\eta,N)$ parameter space as shown in Fig.~\ref{fig_4}.
The line (orange color) separating regions \rom{1} and \rom{2} corresponds to the continuation of the SN bifurcation in the case where the hub sarts in the passive and the periphery in the active state. 
Our reduced system with the two asymmetrically coupled lasers is directed, therefore the passive-active state stability should be considered for the opposite case as well, i.~e. for the hub in the active state and the periphery in the passive. 
This latter bifurcation line (red color) starts at the same coupling strength value as the previous bifurcation line, but has a different behavior and separates the regions \rom{2} and \rom{3}.
Finally, the line (black color) separating regions \rom{3} and \rom{4} corresponds to the continuation of the PB bifurcation that marks the transition from the passive-passive to the active-active state.

These bifurcation lines separate the $(\eta,N)$ parameter space in four distinct regions where the system reaches different steady states.
In the region \rom{1}, any initial condition (periphery active-hub passive, periphery passive-hub active, or periphery passive-hub passive) remains as it is, namely the system is pinned to its initial preparation.
In region \rom{2}, the active periphery drifts the passive hub into the active state.
The same occurs in region \rom{3} where, additionally, the active hub drifts the passive periphery into the active state.
In this region the activation propagates faster from the active periphery towards the passive hub, than from an active hub towards the periphery.
Finally, in region \rom{4} the coupling strength is strong enough even for the periphery passive-hub passive initial condition to jump to the active-active state.
An example of the described dynamical behaviors is shown in he inset of Fig.~\ref{fig_4}, where the evolution of three initial conditions (IC) in the $(\eta,N)$ parameter space is illustrated.
The outer circle represents the periphery of the system, the inner circle represents the hub, while light and dark colors correspond to the passive and active states, respectively.

In close analogy to previous findings in electrochemical bistable networks \cite{Kouvaris2016,  Kouvaris2017} Fig.~\ref{fig_4} shows that the coupling strength required for a transition to occur in the system's dynamics depends on the number of the peripheral nodes.
The line (orange color) separating regions \rom{1} and \rom{2} drops with $\eta$ because as the coupling strength increases, a smaller $N$ size is needed for the periphery to activate the hub (and vice-versa). 
This results in a shift to lower $\eta$ values, of the position of the saddle-node bifurcation when $N$ increases (see Fig.~S3 (b) of the Supplemental Material).
On the other hand, the number of periphery nodes is almost (red line has a tiny slope) independent of the coupling strength required for the hub to activate the periphery (red line and  Fig.~S3 (c) of the Supplemental Material).
Finally, the line separating regions \rom{3} and \rom{4} (black color), which marks the activation of both passive hub and passive periphery, exists for higher values of the coupling strength and, similarly to the orange line, drops with $\eta$ (see Fig.~S3 (d) of the Supplemental Material).

\section{Conclusions}\label{real}
We have shown that star networks of coupled bistable class B lasers support activation spreading from the hub towards the peripheral elements and vice versa.
Interestingly, stationary patterns of activation localized on the hub or the peripheral nodes are also supported, determined by the number of coupled lasers to the central unit, by the coupling strength, and the initial conditions. 
Similar findings were previously reported for electrochemical systems. 
However, the system considered in the current work has been implemented for coupled CO\textsubscript{2} lasers with optoelectronic feedback keeping the bias voltage applied to the modulator constant and by considering the coupling strength as a control parameter. 
After careful numerical calculations, the phases of the central laser and any peripheral unit lock after a very small time interval allowing us to investigate only the steady state of the system.

In a system size-coupling strength diagram we demonstrate four distinct regions indicating different dynamical behavior. 
At weak coupling strengths and small network sizes the initial preparation of the system is pinned and an activation remains stationary and localized either on the peripheral elements or on the hub. 
At weak coupling strengths but larger network sizes an activation can spread only from the periphery towards the hub but not in 
the the opposite direction. Namely an activated periphery turns on the center element but an activated hub cannot drift the periphery to the active state.
This occurs for moderate values of the coupling strength. 
In this third region activation spreads in both directions (with different velocities) and an activated periphery turns on the hub as well as an activated hub can drift the periphery into the active state.
Finally, an activation of the whole system from the passive into the operative region (active state) is shown for strong couplings. 

Despite the obviously different nature of the considered system with the previously studied electrochemical networks, our findings have essential similarities indicating that the network connectivity affects the hosted bistable dynamics in an akin fashion.
The ability to control the spreading or the pinning of an activation, thus the dynamics of the system from the passive into the active states and vice versa, may have multiple technological applications especially in neuromorphic photonics \cite{Pruncal2016, Ferreira2017}, where such tree-like networks can serve for simple hierarchical connectivity structures. 
For future studies, it would be worthwhile to explore if those stationary states can live in the presence of small phase perturbations, due to spontaneous emission or through the detuning of each individual laser cavity length. 
Moreover, it would be interesting to consider the bias regime where the systems exhibits 
oscillations, and instead of the CO\textsubscript{2} laser to study a semiconductor laser diode with a saturable absorber.

\section{Acknowledgments}
This work was supported by the Ministry of Education and Science of the Russian Federation in the framework of the Increase Competitiveness Program of NUST ``MISiS'' (Grant No. K2-2017-006).
N.~E.~K acknowledges financial support by the ``MOVE-IN Louvain'' fellowship, co-funded by the Marie Curie Actions of the European Commission.

\end{document}